\documentclass[usegraphicx]{mn2e}
\usepackage{times}
\newcommand{\plotone}[1]{\includegraphics[width=\columnwidth]{#1}}
\newcommand{\HI}{{\sc H\,i}}
\newcommand{\mJybeam}{mJy beam$^{-1}$}
\newcommand{\msun}{{$M_\odot$}}
\newcommand{\kms}{$\,$km$\,$s$^{-1}$}
\newcommand{\ltsima} {$\; \buildrel < \over \sim \;$}
\newcommand{\gtsima} {$\; \buildrel > \over \sim \;$}
\newcommand{\lta} {\lower.5ex\hbox{\ltsima}}
\newcommand{\gta} {\lower.5ex\hbox{\gtsima}}
\newcommand{\sauron}{{\texttt {SAURON}}}
\newcommand{\miriad}{{\texttt {MIRIAD}}}
\newcommand{\Oiii}{[{\sc O$\,$iii}]}
\newcommand{\Ha}{H$\alpha$}
\newcommand{\Hb}{H$\beta$}

\title[Neutral hydrogen in nearby elliptical and lenticular galaxies
]
       {Neutral hydrogen in nearby elliptical and lenticular galaxies:
 the continuing formation
of early-type galaxies}
\author[R.\ Morganti et al.]
       {R.\ Morganti$^{1,2}$\thanks{E-mail:morganti@astron.nl}, P.T.\
        de Zeeuw$^3$, T.A.\ Oosterloo$^{1,2}$, R.M.\ McDermid$^3$, D.\
        Krajnovi\'c$^4$, \and M.\ Cappellari$^3$, F.\
        Kenn$^{5}$\thanks{ASTRON summer student}, A.\ Weijmans$^3$,
        M. Sarzi$^6$ \\
$^1$ Netherlands Foundation for Research in Astronomy,
    Postbus 2, 7990 AA Dwingeloo, The Netherlands \\
$^2$ Kapteyn Astronomical Institute, University of Groningen
    Postbus 800, 9700 AV Groningen, The Netherlands \\
$^3$ Sterrewacht Leiden, Niels Bohrweg 2, 2333 CA Leiden, The Netherlands\\
$^4$ Denys Wilkinson Building, University of Oxford, Keble Road, Oxford, United
Kingdom \\
$^5$ Argelander-Institut f\"ur Astronomie (AIfA), Universit\"at Bonn,
    Auf dem H\"ugel 71, 53121 Bonn, Germany \\
$^6$ Centre for Astrophysics Research, Science \& Technology Research
    Institute, University of Hertfordshire, Hatfield, United Kingdom\\
}

\pagerange{\pageref{firstpage}--\pageref{lastpage}} \pubyear{2006}

\begin{document}

\maketitle

\label{firstpage}

\begin{abstract}
We present the results of deep Westerbork Synthesis Radio Telescope
observations of neutral hydrogen in 12 nearby elliptical and lenticular
galaxies.  The selected objects come from a representative sample of nearby
galaxies earlier studied at optical wavelengths with the integral-field
spectrograph \sauron.  They are field galaxies, or (in two cases) located in
poor group environments.  We detect \HI\ - both in regular discs as well as in
clouds and tails off-set from the host galaxy - in 70\% of the galaxies.  This
detection rate is much higher than in previous, shallower single-dish surveys,
and is similar to that for the ionised gas.  The results suggest that at faint
detection levels the presence of \HI\ is a relatively common characteristic of
field early-type galaxies,  confirming what was suggested twenty years ago
by Jura based on IRAS observations.  The observed total \HI\ masses
range between a few times $10^6 M_\odot$ to just over $10^9 M_\odot$.  The
presence of regular disc-like structures is a situation as common as \HI\ in
offset clouds and tails around early-type galaxies.  All galaxies where \HI\
isdetected also contain ionised gas, whereas no \HI\ is found around galaxies
without ionised gas. Galaxies with regular \HI\ discs tend to have strong
emission from ionised gas. In these cases, the similar kinematics of the
neutral hydrogen and ionised gas suggest that they form one structure. The
kinematical axis of the stellar component is nearly always misaligned with
respect to that of the gas.  We do not find a clear trend between the presence
of \HI\ and the global age of the stellar population or the global dynamical
characteristics of the galaxies.  More specifically, \HI\ detections are
uniformly spread through the $(V/\sigma, \epsilon)$ diagram. If fast and slow
rotators - galaxies with high and low specific angular momentum - represent
the relics of different formation paths, this does not appear in the presence
and characteristics of the \HI.  Our observations support the idea that gas
accretion is common and does not happen exclusively in peculiar early-type
galaxies.  The links observed between the large-scale gas and the
characteristics on the nuclear scale (e.g., the presence of kinematically
decoupled cores, radio continuum emission etc.), suggest that for the majority
of the cases the gas is acquired through merging, but the lack of correlation
with the stellar population age suggests that smooth, cold accretion could be
an alternative scenario, at least in some galaxies. In either cases, the data
suggest that early-type galaxies continue to build their mass up to the
present.

\date{}
\end{abstract}


\section{Introduction}
\label{sec:introduction}

The currently favoured paradigm for early-type galaxy formation is the
so-called hierarchical formation scenario. It is supported by detailed
N-body and hydrodynamical simulations, which are able to reproduce the
main stellar morphological characteristics and global scaling
relations of early-type galaxies (e.g., De Lucia et al.\
2006\nocite{2006MNRAS.366..499D}). The role and fate of the gas
component, however, is still not well-understood. This is partly due
to the complexity of understanding gaseous processes (star-formation,
energetic feedback and reprocessing), and partly because the gas
content of early-type galaxies is often thought to be insignificant.
Moreover, the parameter space of gas accretion is large, ranging from
merging of two large, equal mass gas-rich objects to infall of a tiny
gas-rich companion, and perhaps so-called cold accretion, the slow but
long-lasting infall of primordial gas (e.g., Keres et al.\
2005\nocite{2005MNRAS.363....2K}).

Recent observations of early-type galaxies show that gas is clearly present in
these objects, and that gas processes may play a more important role in
shaping the stellar properties than previously thought. In a galaxy merger event,
the amount of gas involved can have a profound effect on the merger remnant,
with gas-rich events leading to more disc-like objects (e.g., Bekki \& Shioya
1997\nocite{1997ApJ...478L..17B}; but see Burkert \& Naab
2005\nocite{2005MNRAS.363..597B}). This may explain the `discy' isophote
distortions present in many early-type galaxies, and it has been suggested as an
explanation of the apparent dichotomy of fast- and slow-rotating galaxies
(i.e.  galaxies with high and low specific angular momentum, Bender, Burstein
\& Faber 1992). Dynamically distinct stellar sub-components are often found in
early-type galaxies, and taken as evidence for formation via
merging. Connecting a significant star-formation event with such a merger has
given mixed results, but there are clearly cases where sub-components of the
galaxy are both chemically and kinematically distinct (McDermid et al.\ 2006),
strongly suggesting that external gas has entered the system.

In order to make progress on these issues, high-quality observations of
the gas content of early-type galaxies are crucial to allow detailed
comparisons with the stellar properties. In the optical, various
studies in the recent past have explored the characteristics of the
ionised gas in early-type galaxies (Phillips et al.\
1986\nocite{1986AJ.....91.1062P}; Buson et al.\
1993\nocite{A&A...280..409B}; Goudfrooij et al.\
1994\nocite{1994A&AS..105..341G}) and found the presence of gas with
complex kinematics (e.g., counter-rotating with respect to the stellar
component, Bertola et al.\ 1992\nocite{1992ApJ...401L..79B}). However,
so far the kinematics and ionisation of the gas in early-type galaxies
have been studied mostly through long-slit observations, usually along
one or two position angles, which limit the correct
determination of the morphology and dynamical structure of the
ionised gas. The recent
systematic survey based on observations with the panoramic
integral-field spectrograph \sauron\ shows that these objects display
a variety of line-strength distributions and kinematic structures
which is richer than often assumed (Bacon et al.\
2001\nocite{2001MNRAS.326...23B}; de Zeeuw et al.\
2002\nocite{2002MNRAS.329..513D}). The survey includes 48
representative nearby early-type galaxies classified as E or S0 in the
RC3 (de Vaucouleurs et al.\ 1991\nocite{1991trcb.book.....D}).
Many examples of minor axis rotation, decoupled cores, embedded
metal-rich stellar discs, as well as non-axisymmetric and
counter-rotating gaseous discs, have been found (Emsellem et al.\
2004\nocite{2004MNRAS.352..721E}; Sarzi et al.\
2006\nocite{2006MNRAS.366.1151S}; Kuntschner et al.\
2006\nocite{keta06}).

At radio wavelengths, our knowledge about the neutral hydrogen content of
early-type galaxies is also changing. This is partly due to the growing number
of cases where \HI\ has been imaged --- instead of using only single-dish data
--- and information about the morphology and the detailed kinematics of the
gas is now available.  Many \HI-rich early-type galaxies are now known
(e.g. Schiminovich et al.\ 1995\nocite{1995ApJ...444L..77S}; van Gorkom \&
Schiminovich 1997\nocite{1997ASPC..116..310V}; Morganti et al.\
1997\nocite{1997AJ....113..937M}; Sadler et al.\
2000\nocite{2000AJ....119.1180S}; Balcells et al.\
2001\nocite{2001AJ....122.1758B}; Oosterloo et al.\
2002\nocite{2002AJ....123..729O} and refs therein; Oosterloo et al.\
2004\nocite{2004IAUS..217..486O}, 2005). The large amount of neutral hydrogen
detected around some of these galaxies (up to more than $10^{10} M_\odot$) is
often distributed in huge (up to 200 kpc in size) regularly rotating discs or
rings. The flat rotation curves of the discs indicate the existence of large
halos of dark matter.  Given their regular appearance, these discs must be
relatively old (several $\times 10^9$ yr).  Although an external origin of the
\HI\ has been suggested already in several earlier studies (e.g. Knapp et al.\
1985\nocite{1985AJ.....90..454K}), the imaging of the kinematics of such
structures opens the possibility of studying in detail how these galaxies have
formed. In particular, although major mergers and small-companion accretions
are clearly at the origin of some of the
\HI\ structures observed (e.g. Serra et al.\
2006\nocite{2006astro.ph..2621S}),
recent work has shown that smooth cold gas
accretion (e.g., Macci\'o, Moore \& Stadel 2006\nocite{2006ApJ...636..25M})
can also play an important role and should, therefore, be taken into account.

The shallow \HI\ surveys available so far are, however, able to
study only the most extreme \HI-rich early-type galaxies. Much deeper
observations are needed to explore the complete \HI\ mass distribution
of early-type galaxies.  Furthermore, the study of \HI\ in these
systems has lacked the important combination of having both the \HI\
data and multi-slit or integral-field optical spectroscopy available
for a significant number of objects. For these reasons, we have
performed deep \HI\ observations, using the recently upgraded
Westerbork Synthesis Radio Telescope (WSRT), of a sub-sample of E and
S0 galaxies in the \sauron\ representative survey.  We present the
results here. We describe the sample selection and the WSRT
observations in Section~\ref{sec:observations}. In
Section~\ref{sec:results} we discuss the \HI\ maps, and we compare
with earlier \HI\ surveys in Section~\ref{sec:comparison_hi}. A
discussion of the relation between the presence of \HI\ and of a radio
loud AGN is given in Section~\ref{sec:comparison_agn}. We investigate
the relation with the characteristics of the stellar component and the
ionised gas in Section~\ref{sec:comparison_sauron}, and comment on the
origin of the neutral gas in Section~\ref{sec:origin}.  We summarise
our conclusions in Section~\ref{sec:conclusions}. In
Appendix~\ref{appendix:a} we report the serendipitous discovery of a
megamaser in the field of NGC 4150.

\section{Sample and \HI\ WSRT observations}
\label{sec:observations}

The \sauron\ sample contains 24 galaxies classified as E in the RC3, and
another 24 classified as S0. They are divided equally between so-called
`field' and `cluster' environments, and cover a factor 50 in total luminosity
and the full range of ellipticity (de Zeeuw et al.\
2002\nocite{2002MNRAS.329..513D}). We selected the 12 E and S0 objects with
declination $\delta > 23^\circ$, in order to have good spatial resolution with
the WSRT.  The majority of the selected galaxies are genuine field galaxies,
with two cases (NGC~4150 and NGC~4278) are located in poor group
environments. None of them reside in a dense cluster environment.

The specifics of the WSRT observations are listed in
Table~\ref{tab:table1}. The observations were made using a band of 20~MHz
(corresponding to $\sim$4000 \kms), centred on the frequency of the redshifted
\HI, and sampled with 1024 channels. One object, NGC\,2685, had already been 
observed with a similar setup by J\'ozsa et al.\
(2004a\nocite{2004bdmh.confE.108J}) and J\'ozsa (2006\nocite{joszaphd}). We
did not re-observe this galaxy but refer to their results.

The calibration and analysis were done using the \miriad\ package. The
data cubes were constructed with a robust-Briggs (1995) weighting
equal to 0, or with natural weighting for the faintest cases. The
cubes were made by averaging channels in groups of two, followed by
Hanning smoothing so that a velocity resolution of 16 \kms\ was
obtained. This was done to optimise sensitivity. The r.m.s.\ noise and
restoring beam sizes of each cube are given in Table~\ref{tab:table1}.

As a by-product of the observations, the line-free channels were used
to obtain an image of the radio continuum of each galaxy. The
continuum images were made with uniform weighting. The r.m.s.\ noise
and beam of these images are also given in Table~\ref{tab:table1}.
Radio continuum emission was not detected in four of the objects, in
which cases only upper  limits are determined. All the detected
continuum sources are unresolved. The peak flux and power of the
continuum, or the $3\sigma$ upper limits, are given in
Table~\ref{tab:table2}.

\begin{table*}
\tabcolsep=3pt
\begin{tabular}{ccccccccccc}
\hline\hline
NGC  &Type   & $V_{\rm centr}$ & D &pc/$^{\prime\prime}$ &Date & Int.Time & Beam & Noise \HI\ & Noise Cont. & \HI\ contours \\
     &     &       \kms\   & Mpc  &                      &     &   h      & $^{\prime\prime}\times ^{\prime\prime} (^\circ)$
                                           & \mJybeam\  & \mJybeam\   & $10^{19}$ cm$^{-2}$       \\
(1)  & (2)   &  (3)  & (4)  & (5)          & (6) & (7)      & (8)  & (9)
& (10)    & (11)      \\
\hline
1023 &  S0  & 614  &11.4 &   55   &$\!\!\!$ 12-17/10/04 &$4\!\times\!12$ &$21\!\times\!14$(1) &0.26 &0.026 &(2.5),5,10,25,50   \\
2549 &  S0  & 1069 &12.6 &   61   & 02/01/04    &12              & $60\!\times\!55$(0) &0.41 &0.034 & --                     \\
2685 &  S0  & 883  &15.7  &   76   &$\!\!\!$ 12/02-01/03$^{a}$ &$4\!\times\!12$  &$28\!\times\!25$(10)&0.24 &-- &(5),10,25,50,100\\
2768 &  E   & 1359 &22.4  &  109   &03/01/04     &12   &$33\!\times\!33$(-84) &0.47 &0.034           &1,2.5,5,10,25           \\
3414 &  S0  & 1472 &25.2  &  122   & 2/02/04     &$4\!\times\!12$ &$45\!\times\!33$(9)  &0.27 &0.026 &1,2.5,5,10,50,100       \\
4150 &  S0  & 219  &13.7  &   67   &31/01/04     &$4\!\times\!12$ &$41\!\times\! 34$(17)&0.30 &0.021 &1,2.5                   \\
4278 &  E   & 631  &16.1  &   78   &04/02/04     &$4\!\times\!12$ &$28\!\times\!14$(11) &0.23 &0.037 &1,2.5,5,10,25           \\
5198 &  E   & 2531 &39.6  &  192   &01/05/04     &$2\!\times\!12$ &$37\!\times\!35$(13) &0.37 &0.026 &$\!\!\!$2.5,5,10,25,50,100 \\
5308 &  S0  & 1985 &32.8  &  159   &22/04/04     &12              &$35\!\times\!35$(0)  &0.58 &0.080 & --                     \\
5982 &  E   & 2935 &45.7  &  222   &11/09/04     &12              &$36\!\times\!36$(79) &0.56 &0.043 &5,10 \\
7332 &  S0  & 1206 &23.0  &  112   &07/09/04     &18       &$46\!\times\!31$(6)  &0.40 &0.043 &$\!\!\!\!\!$1,2.5,5,10,25,50,100 \\
7457 &  S0  & 845  &13.2  &   64   &31/08/04     &12              &$39\!\times\!32$(15) &0.38 &0.038 & --                     \\
\hline
\end{tabular}
\caption{Summary of observations and properties of the galaxies in the sample.
(1) Galaxy identifier.  (2) Systemic velocity at which we centred the
\HI\ observation band. (3) Hubble type (NED). (4) Galaxy distance from the SBF
measurements of Tonry et al.\ (2001).  Four galaxies (NGC~2685, NGC~5198,
NGC~5308 and NGC~5982) do not have SBF distances.  In those cases we used
redshift distances from LEDA.  (5) Linear scale. (6) Date of observation. (7)
Integration time in hours. (8) Beam. (9) Noise level in the \HI\ cube. (10)
Noise level of the continuum image. (11) Contour levels of the total
intensity images shown in Fig.~\ref{fig:himaps}. Note: $(a)$ Data taken by G.\
J\'ozsa and presented in J\'ozsa et al.\ (2004a) and J\'ozsa (2006).\label{tab:table1}}
\end{table*}

\begin{table*}
\tabcolsep=3pt
\begin{tabular}{cccccccccc}
\hline\hline
NGC    &Type& K$_T$      &$M_{\rm HI}$       &$M_{\rm HI}/L_B$ & $M_{\rm HI}/L_K$ &  Diam \HI\ & S$_{\rm 1.4GHz}$ & log$P_{\rm 1.4GHz}$& \HI\ references \\
       &    &  mag       & $M_\odot$         &           &          &kpc & mJy     &  W/Hz   &      \\
(1)    &(2) &  (3)       & (4)               & (5)       & (6)      & (7)& (8)    & (9)    & (10) \\
\hline
 1023  & S0 &  6.24      & $2.1\times 10^9$  &    0.046  &    0.025 & 92 &    0.4  &  18.7   & a  \\
 2549  & S0 &  8.05      & $<2.0\times 10^6$ &$<0.00043$ &$<0.00010$& -- &   $<0.1$&$<18.3$  &    \\
 2685  & S0 &  8.35      & $1.8\times 10^9$  &     0.27  &    0.078 & 41 &  2$^*$  &  19.8   & b,c,d \\
 2768  & E  &  7.00      & $1.7\times 10^8$  &   0.0038  &   0.0010 & 60 &   10.9  &  20.8   &    \\
 3414  & S0 &  7.98      & $1.6\times 10^8$  &   0.0096  &   0.0019 & 26 &    5.0  &  20.6   &    \\
 4150  & S0 &  8.99      & $2.5\times 10^6$  &  0.00078  &  0.00025 &  4 &    0.8  &  19.2   &    \\
 4278  & E  &  7.18      & $6.9\times 10^8$  &    0.039  &   0.0097 & 37 &  336.5  &  22.0   & e,f  \\
 5198  & E  &  8.90      & $6.8\times 10^8$  &    0.030  &   0.0077 & 70 &    2.4  &  20.6   &    \\
 5308  & S0 &  8.36      & $<1.5\times 10^7$ &$<0.00064$ &$<0.00015$& -- &  $<0.24$&$<19.5$  &    \\
 5982  & E  &  8.15      & $3.4\times 10^7$  &  0.00060  &  0.00014 & -- &    0.5  &  20.1   &    \\
 7332  & S0 &  8.01      & $6.0\times 10^6$  &  0.00038  &  0.00009 & -- &  $<0.13$&$<18.9$  &    \\
 7457  & S0 &  8.19      & $<2.0\times 10^6$ &$<0.00032$ &$<0.00010$& -- &  $<0.11$&$<18.5$  &    \\
\hline
\end{tabular}

\caption{Further properties of the galaxies and measurements based on our
radio observations.  (1) Galaxy identifier.  (2) Hubble type (NED).  (3) Total
$K$-band apparent magnitude from the 2MASS Extended Source Catalogue (Jarrett
et al.\ 2000).  (4) Total mass in \HI.  (5) Ratio of total mass in \HI\ and
the absolute $B$-band luminosity $L_B$.  (6) Ratio of total mass in \HI\ and
the absolute $K$-band luminosity $L_K$.  (7) Diameter of the \HI\
distribution.  (8) Continuum flux (or $3\sigma$ upper limits) at 1.  4 GHz. 
(9) Total radio power at 1.4 GHz.  (10) References: $(a)$ Sancisi et al.\
(1984); $(b)$ Shane (1980); $(c)$ J\'ozsa et al.\ (2004a); $(d)$J\'ozsa
(2006); $(e)$ Raimond et al.\ (1981); $(f)$ Lees (1994); Note: * radio
continuum from NVSS.\label{tab:table2}} \end{table*}

\section{Results}
\label{sec:results}

We detect \HI\ in emission in eight (possibly nine, NGC\,7332 is an
unclear case, see Section~\ref{sec:results_offset}) of the 12 galaxies
observed. Three of the detected objects (NGC\,1023, NGC\,2685 and
NGC\,4278) were already known to have \HI\ from previous observations
(respectively: Sancisi et al.\ 1984\nocite{1984MNRAS.210..497S}; Shane
1980\nocite{1980A&A....82..314S}; Raimond et al.\
1981\nocite{1981ApJ...246..708R} and Lees
1994\nocite{1994mtia.conf..432L}). Table~\ref{tab:table2} summarises
the \HI\ morphology, mass and size for every detected object. Three
galaxies (NGC\,2549, NGC\,5308, NGC\,7457) are not detected in \HI. In
these cases the \HI\ mass limits range from a few times $10^6 M_\odot$
to at most $10^7 M_\odot$.

\begin{figure*}

\includegraphics[width=14cm]{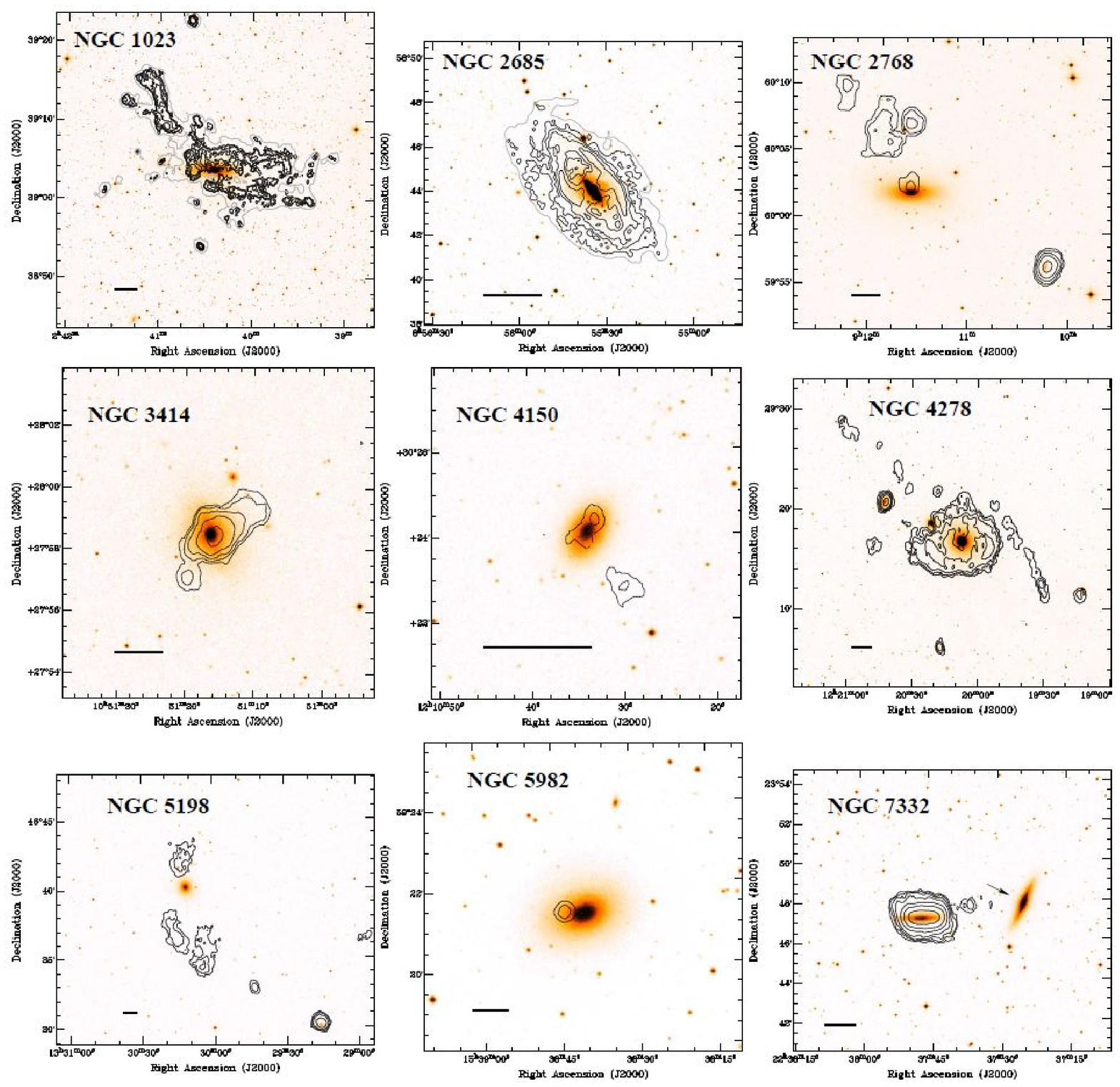}

\caption{Total \HI\ intensity images (contours) superimposed onto  Digital
Sky Survey optical images of the detected galaxies. Grey contours in the
figures of NGC\,1023 and NGC\,2685 correspond to the emission detected from
lower resolution data. The contour levels are given in Table~\ref{tab:table1}
with the grey contour given in parenthesis.  The 10 kpc linear scale is given
in every plot. NGC~7332 is indicated with an arrow.
Further information on the \HI\ in NGC\,2685, can be found in J\'ozsa et
al.\ (2004a) and J\'ozsa (2006).\label{fig:himaps}}
\end{figure*}

The measured \HI\ masses (see Table~\ref{tab:table2}) range between a few
times $10^6 M_\odot$ to just over $10^9 M_\odot$.  The value for the gas
content ($M_{\rm HI}/L_B$) ranges
from $< 0.0003 M_\odot/L_\odot$, to as high as $0.3 M_\odot/L_\odot$ in the
case of NGC\,2685. The latter represents a value at the gas-rich end of the
distribution characteristic of early-type galaxies (Knapp et al.\
1985\nocite{1985AJ.....90..454K}) and is comparable to that of normal spiral
galaxies.

Fig.~\ref{fig:himaps} shows the \HI\ total intensity images. The sizes of the \HI\
structures vary between 30 and 90~kpc with the exception of the tiny
\HI\ disc detected in NGC\,4150 that is only 4~kpc in diameter and the
small cloud in NGC~5982. The typical peak column density is at most a
few times 10$^{20}$ cm$^{-2}$ (see contour levels in Fig.~\ref{fig:himaps}). As
already found from the \HI\ observations of other early-type galaxies
(see, e.g., van Driel \& van Woerden 1991\nocite{1991A&A...243...71V};
Morganti et al.\ 1997\nocite{1997AJ....113..937M}; Serra et al.\
2006\nocite{2006astro.ph..2621S}) these values of the column density
are lower than the critical surface density for star formation
proposed by Kennicutt (1989\nocite{1989ApJ...344..685K}). Although
this result excludes the presence of widespread star formation
activity, local small regions of star formation can still be present.

Most of the continuum sources associated with the observed galaxies
are too weak to be used to detect \HI\ in absorption. The only
exception is NGC\,4278, where nevertheless no \HI\ absorption was
found. Given the low spatial resolution of our observations, \HI\
absorption, even if present, is likely filled up with the \HI\ in
emission present in this galaxy.

\begin{figure*}

\includegraphics[width=14cm]{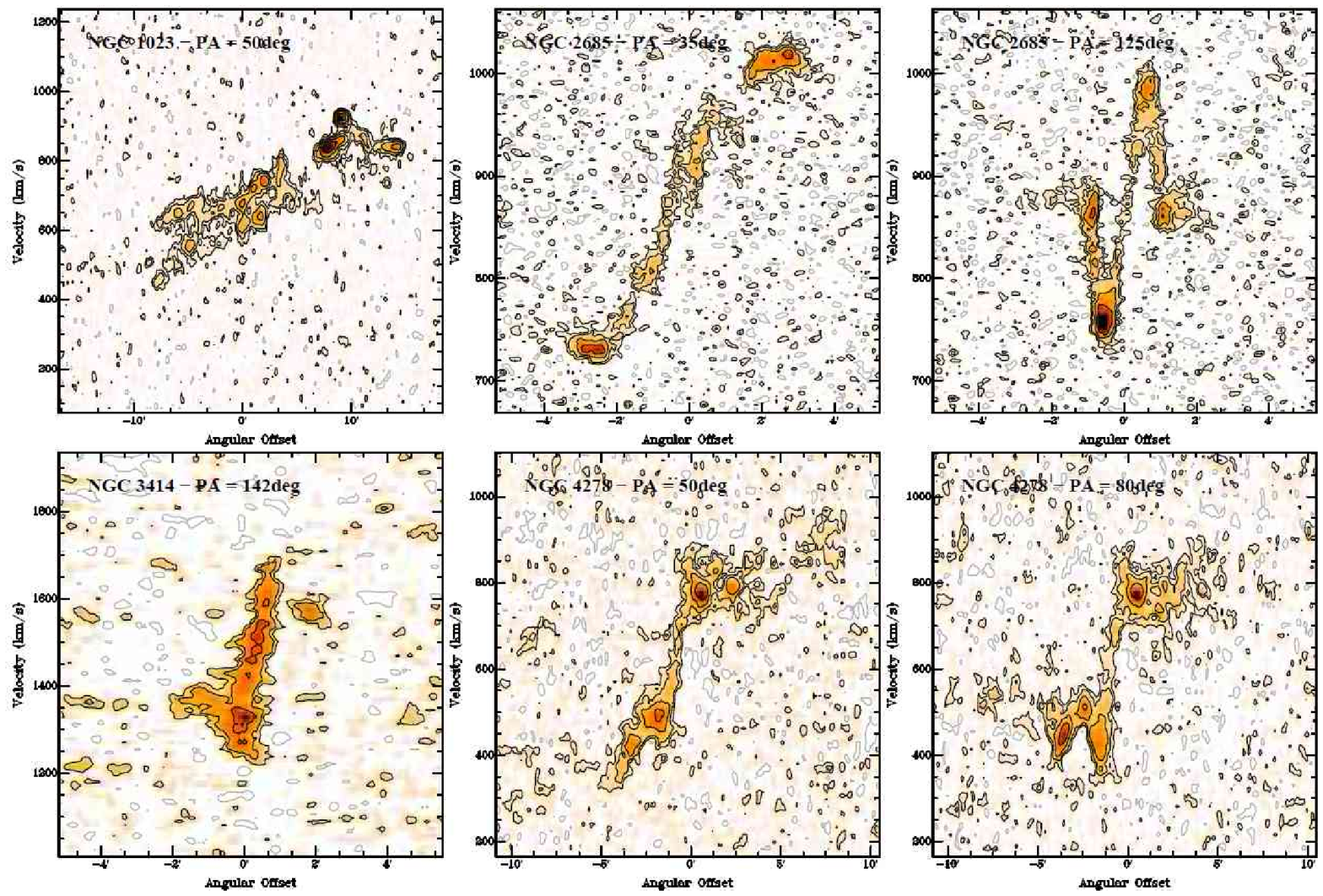}
\caption{Position-velocity plots of some of the detected galaxies. Contours
levels are: NGC~1023; $-0.32, 0.32:0.1 \times 2$ mJy beam$^{-1}$;
NGC~2685: $-0.54, 0.54:0.1 \times 2$ mJy beam$^{-1}$; NGC~3414:
$-0.25, 0.25:0.01 \times 2$ mJy beam$^{-1}$; NGC~4278: $-0.21,
0.21:0.01\times 2$ mJy beam$^{-1}$. \label{fig:pvplots} }
\end{figure*}

We divide the observed \HI\ structures in three main groups: 1) regularly
rotating, disc-like \HI\ emission, 2) offset clouds or tails and 3) complex
distribution.  The presence of regular disc-like structures is as common as
\HI\ in offset clouds and tails around galaxies. Fig.~\ref{fig:pvplots} shows
examples of position-velocity plots, along one of the main axes, for the most
interesting cases. Below, we summarise the \HI\ characteristics for each
object, together with relevant information obtained from the \sauron\
observations. For a more detailed description of the optical characteristics
obtained from \sauron\ we refer to the original papers: Emsellem et al.\
(2004\nocite{2004MNRAS.352..721E}), Sarzi et al.\
(2006\nocite{2006MNRAS.366.1151S}) and Kuntschner et al.\
(2006\nocite{keta06}).

\subsection{Regular \HI\ structures}
\label{sec:results_regular}

In four galaxies (NGC\,2685, NGC\,3414, NGC\,4150 and NGC\,4278), the
\HI\ appears to be distributed in a relatively regularly rotating
structure.

NGC\,2685, also known as the Spindle galaxy, shows strong, helix-like
dust extinction on the North-Eastern side and it is a classical
polar-ring (Whitmore et al.\ 1990\nocite{1990AJ....100.1489W}).  The
kinematics of the \HI\ clearly change with radius.  In the inner part
the kinematical major axis is perpendicular to the photometric major
axis, while at large radius, the kinematical major axis is aligned
with the photometric major axis. The two position-velocity diagrams in
Fig.~\ref{fig:pvplots} show this very clearly. A tilted-ring analysis (J\'ozsa et al.\
2004a\nocite{2004bdmh.confE.108J}; J\'osza 2006) indicates that the
\HI\ is actually one single structure that is polar in the inner
regions while at a radius of about 1$^\prime$ it warps about
$90^\circ$ to become co-planar with the optical galaxy.  J\'ozsa et
al.\ (2004a\nocite{2004bdmh.confE.108J}) suggest that the \HI\ could
be a disc that forms from a single accretion event under the influence
of a tumbling triaxial halo (Peletier \& Christodoulou
1993\nocite{1993AJ....105.1378P}).
The \sauron\ data show the presence of ionised gas (\Hb\ and \Oiii),
mostly concentrated along the photometric minor axis.  The kinematics
of the ionised gas and of the stars is shown in Fig.~\ref{fig:neutral_ionised}. This figure
clearly shows that the ionised gas has the same kinematics as the very
inner regions in the \HI\ data.  The stellar populations are
moderately young and have solar metallicity.  CO (J=2-1 and
J=1-0) emission was detected by Schinnerer \& Scoville
(2002\nocite{2002ApJ...577L.103S}). They find four molecular cloud
associations in the Western and Eastern regions of the galaxy, close
to the brightest \Ha\ and \HI\ peaks in the polar ring. The CO and
\HI\ line velocities agree, which indicates that the CO emission also
originates from the polar ring.

NGC~3414 is another interesting object. A rotating structure of
$\sim$$3.5^\prime$ ($\sim$29 kpc) and $\sim$$10^8 M_\odot$ of neutral hydrogen
is observed. The rotation axis of the \HI\ is misaligned with the photometric
axis by $\sim$$44^\circ \pm 5^\circ$. The \HI\ position-velocity diagram
(Fig.~\ref{fig:pvplots}) suggests the presence of two kinematical components,
a fast (at least in projection) inner one and a slower, or possibly more
face-on, extended outer structure. In the inner region, observed with \sauron, the stellar kinematics
shows a decoupled core (KDC).  The ionised gas has a complex morphology
showing a smoothly twisting velocity field such that the rotation axis aligns
with the photometric major axis at large radii, but it almost aligns with the
KDC within the central $5^{\prime\prime}$ radius (see also
Fig.~\ref{fig:neutral_ionised}). The rotation of the \HI\ lines up very well
with that of the outer ionised gas.  The stellar populations is old
and has solar metallicity.

NGC\,4150 shows the faintest and smallest rotating \HI\ disc detected,
with only a few $\times 10^6 M_\odot$ of neutral hydrogen. The size of
the \HI\ disc is about 1$^\prime$ ($\sim$4 kpc). A cloud of low column
density \HI\ is also observed about $1^{\prime}$ from the galaxy,
without an obvious optical  counterpart.  NGC~4150 is also detected in
CO (Welch \& Sage 2003\nocite{2003ApJ...584..260W}; Leroy et al.\
2005) and the estimated molecular gas content is $\sim$$3.0
\times 10^7 M_\odot$.  From the \sauron\ data, a small stellar KDC is
observed in the central 2$^{\prime\prime}$. The ionised gas follows
the outer stellar kinematics, with a possible central disc.  The
stellar population is globally rather young with a
strong contribution of young stars in the central few arcseconds.

NGC\,4278 was known to have an extended regular disc of \HI\ (Raimond et al.\
1981\nocite{1981ApJ...246..708R}; Lees 1994\nocite{1994mtia.conf..432L}), but
our data show this disc and its kinematics in much more detail. The new data
also show that two faint tail-like structures exist at large radius (see
Fig.~\ref{fig:himaps}). The kinematics of the \HI\ is regular. The
position-velocity diagram taken along position angle (PA) 80$^\circ$ 
(Fig.~\ref{fig:pvplots}) shows large modulations of the velocities indicating
that large deviations from a flat disc in circular rotation occur.  The
\sauron\ data show strong ionised gas, with a large-scale twisted rotation
field, rotating in the same sense as the stars, but misaligned by $\sim$20--70
degrees (see also Fig.~\ref{fig:neutral_ionised}). Also in this galaxy, the
kinematics of the \HI\ and the ionised gas match very well, and both are
misaligned with the stellar kinematics and the photometric axes. The stellar
population is old and has near-solar metallicity.

\subsection{Neutral hydrogen off-set from the galaxy}
\label{sec:results_offset}

In three objects (NGC\,2768, NGC\,5198, NGC\,5982) most of the \HI\ is
found in clouds/tails offset from the centre of the optical galaxy
(see Fig.~\ref{fig:himaps}).  These \HI\ features do not have an obvious
stellar counterpart.  They have velocities that are very close (within 300
\kms) to the systemic velocity of the target galaxy. They are,
therefore, likely physically associated with the galaxy.

In NGC\,2768, most of the \HI\ is detected in a tail-like system at
about 16 kpc NE from the centre. However, some faint \HI\ emission is
also present inside the optical boundaries.  The kinematics suggest
that these two features are related. The neutral hydrogen in the cloud
shows velocities that range from 1430 to 1510 \kms, slightly
redshifted compared to the systemic velocity of the galaxy (1373
\kms). This velocity matches that of the ionised gas on this side of
the galaxy.  Wiklind et al.\ (1995\nocite{1995A&A...297..643W})
detected this galaxy in CO with IRAM, and infer a molecular hydrogen
mass of $\sim$$2 \times 10^7$ \msun. The ionised gas detected by
\sauron\ rotates around the major axis. The rotation is perpendicular
to that of the stars, and the stellar population is old.

In NGC\,5198 we detect \HI\ both $\sim$$2^\prime$ ($\sim$23 kpc) north
and $\sim$$4^\prime$ ($\sim$46 kpc) south of the galaxy. The current
data are not deep enough to see whether these are actually part of a
single, large gaseous structure. However it is interesting that the
systemic velocity of the galaxy is exactly in between that of the two
\HI\ clouds.  The \sauron\ data show a central stellar KDC misaligned
with the outer rotation.  The ionised gas is mostly detected in the
central few arcseconds. However, the ionised gas does also extend to the
North, and hence could be related to the \HI\ structure. The ionized gas
occupies a disc-like structure which is perpendicular to the central
KDC, and counter-rotates with respect to the outer rotation. The
stellar population is old.

In NGC\,5982, a very small blob of \HI\ appears just offset ($\sim$6
kpc) from the galaxy centre with a velocity of $\sim$2830 \kms, about
200 \kms\ lower than the systemic velocity. The \sauron\ data show a
small amount of ionised gas and an apparent KDC. The central gas
rotates in the same sense as the KDC.  The stellar population is old.

For NGC\,7332, an \HI\ cloud is detected {\sl between} this galaxy and
the \HI-rich companion galaxy NGC\,7339. The cloud of neutral hydrogen
is located about $3^\prime$ ($\sim$14 kpc) from the centre of
NGC\,7332 (see Fig.~\ref{fig:himaps}). The velocity of the \HI\ cloud ($\sim$$1250$
\kms) is roughly similar to that of the western side of the \HI\ disc
of the companion galaxy NGC\,7339 but is also similar to the velocities of the
ionised gas measured in the eastern side of the NGC\,7332.  The velocity of
the \HI\ cloud is, however, quite different from the anomalous high-velocity
($\sim$$300$~\kms\/ around the systemic velocity) ionised gas detected on the
southern side of the galaxy centre in NGC\,7332 (Falc\'on--Barroso et al.\
2004\nocite{2004MNRAS.350...35F}). The neutral hydrogen is likely to be a
signature of some interaction between the two galaxies. This interpretation is
consistent with evidence found from the \sauron\ data and from Howell
(2006). The stellar kinematics derived by \sauron\ show regular rotation, with
a small KDC within the central 3 arcseconds. The ionised gas is
complex, with some regularly rotating structure in \Oiii\ within the central
10 arcsec, which seems to lead onto a larger-scale structure rotating around
the long axis (misaligned by 90 degrees to stellar rotation). The stellar
population is young. NGC~7332 has been also detected in CO emission by
Welch \& Sage (2003\nocite{2003ApJ...584..260W}) and they estimate an upper
limit to the molecular hydrogen content of $4.2 \times 10^7$
\msun.

\subsection{Complex \HI\ kinematics}
\label{sec:results_complex}

The galaxy NGC\,1023 has a very extended and complex neutral hydrogen
distribution. This \HI\ has been studied in detail by Sancisi et al.\
(1984\nocite{1984MNRAS.210..497S}). Our observations are a factor of
five deeper and have much better spatial and velocity resolution.  The
morphology of the \HI\ (see Fig.~\ref{fig:himaps}), however, does not reveal any
major surprises compared to the study of Sancisi et al.\
(1984\nocite{1984MNRAS.210..497S}). On large scales, the \HI\ appears,
to first order, to be rotating around the galaxy. However, double-peaked
profiles are observed throughout the main body of \HI\ (see Fig.~\ref{fig:pvplots}),
showing that large, systematic deviations from circular rotation
occur, while small components with opposite velocity gradients exist.
Sancisi et al.\ (1984\nocite{1984MNRAS.210..497S}) describe the
distribution of \HI\ as reminiscent of the tails and bridges found in
interacting galaxies and this object may represent an example of a
merger in an intermediate stage when the gas is still in the process
of settling. Indeed, one of the brightest clouds in the \HI\
distribution coincides with a faint optical companion. The
\HI\ is very clumpy, quite different in character from what we
observe in the other galaxies.  Welch \& Sage
(2003\nocite{2003ApJ...584..260W}) estimate an upper limit to the
molecular gas mass of $4.9 \times 10^7$ \msun.  From the \sauron\
data, the stellar kinematics of NGC\,1023 show a small but steady
twist in the rotation axis across the \sauron\ field, consistent with the
presence of a large stellar bar.  The ionised gas is patchy, although
seems to rotate in a similar sense as stars and the \HI. No obvious
dust features are observed. Finally, the stellar population is old
with a slightly super-solar metallicity.\looseness=-2

\subsection{Undetected \HI\ galaxies}
\label{sec:results_undetected}

For completeness, we briefly summarise the properties (in particular the
optical properties derived from the \sauron\ data) of the three
galaxies undetected in \HI.

NGC\,2549: the ionised gas shows a disc-like structure in the central few
arcseconds which is aligned with the stars, but has an irregular distribution
at larger radii, showing perpendicular rotation to the stars and a filamentary
structure. The stellar populations are of intermediate age and have
super-solar metallicity.

NGC\,5308 also has very discy stellar kinematics, with an additional
thin, edge-on disc in the central 5 arcseconds. No emission-line
gas was detected with \sauron. The stellar populations are old with solar metallicity.

NGC\,7457 is a regularly rotating flattened object, with a small KDC
inside the central 3 arcseconds. Not much ionised gas was
detected; it has irregular structure at large radius, and a compact
central rotating component, seemingly aligned with the KDC. The
stellar populations in this galaxy are quite young and have
solar metallicity. Observations of CO are reported by Welch
\& Sage (2003\nocite{2003ApJ...584..260W}), which resulted in an
estimated upper limit of $3.3 \times 10^7 $\msun\ for molecular
hydrogen.

\section{Comparison with other \HI\ surveys}
\label{sec:comparison_hi}

The survey of \HI\ in emission presented here represents one of the
few available so far for a representative sample of regular early-type
galaxies, where both imaging and kinematics of the gas is obtained and
where also detailed optical spectral maps are available.  Most of the
previous \HI\ surveys have used single-dish observations (see e.g.\
Huchtmeier et al.\ 1983\nocite{1983gcho.book.....H}; Knapp et al.\
1985\nocite{1985AJ.....90..454K}). A summary of the detection rates is
given in van Gorkom \& Schiminovich (1997\nocite{1997ASPC..116..310V})
and Knapp (1999\nocite{1999ASPC..163..142K}) as well as in other
papers (Roberts et al.\ 1991\nocite{1991ApJS...75..751R}; Bregman et
al.\ 1992\nocite{1992ApJ...387..484B}; Huchtmeier et al.\
1995\nocite{1995A&A...300..675H}). Although the morphological
classification is always a source of uncertainty, these authors
estimate the detection rate for field E and S0 galaxies to be 5 and
20\% respectively, whilst having a galaxy classified as peculiar
greatly enhances the chance of it being detected in \HI, with a
detection rate of 45\% for Pec E and S0 galaxies.

The surprisingly high detection rate of our observations is likely to
be due to a combination of the depth that we have reached and the fact
that, unlike single-dish surveys, we have imaged the distribution of
\HI\ in every galaxy, therefore being able to associate to our target
galaxies even small clouds of \HI\ (such as in the case of NGC\,5982).
The results of our deep \HI\ imaging survey indicate that, at faint
detection levels, {\sl the presence of \HI\ could be a relatively
common characteristic of many early-type galaxies in the field}.  This
is an important result, although it should be verified by a larger sample.

Jura (1986) reported a high detection rate of E and S0 galaxies in the
IRAS 60 and 100 micron bands, and deduced the presence of cold dust that would
presumably be accompanied by cold gas. He estimated that about one-third of
the E/S0's (and maybe more) contain of the order of $10^8 M_\odot$ in \HI. Our
\HI\ masses agree nicely with his prediction, but the detection rate is even
higher than he suggested.

The morphologies of the neutral hydrogen found in our survey can be compared
with the results from, e.g., van Driel \& van Woerden
(1991\nocite{1991A&A...243...71V}). These authors have studied a sample
selected from gas-rich S0 galaxies.  These galaxies have neutral hydrogen
distributed mainly in inner or outer \HI\ rings. The kinematics can be
described mainly by circular rotation --- with cases of warped distributions
--- and flat rotation curves. Compared to these results, our deeper survey has
shown the presence of other cases where the \HI\ is offset compared to the
optical galaxy suggesting a more complex picture of the \HI\ properties of S0
galaxies.  In agreement with the results of van Driel
\& van Woerden (1991\nocite{1991A&A...243...71V}), the average \HI\
surface density found is only $\sim$$1 M_\odot$ pc$^{-2}$, too low to
sustain large-scale star formation.

From the selection of early-type galaxies from the cross-correlation of the
\HI\ Parkes All-Sky Survey (HIPASS, Barnes et al.\
2001\nocite{2001MNRAS.322..486B}) and the RC3 catalogue, together with imaging
follow-up using ATCA (Sadler et al.\ 2000\nocite{2000AJ....119.1180S};
Oosterloo et al.\ 2004\nocite{2004IAUS..217..486O}, 2005\nocite{oeta05}) it
has been shown that between 5 and 10\% of early-type galaxies are extremely
gas rich (with \HI\ masses well above $ 10^9 M_\odot$ and value of $-1<
\log(M_{\rm HI}/L_B) < 0$). In about 70\% of these \HI-rich
galaxies, the neutral hydrogen appears to be distributed in extremely
large (up to 200 kpc in diameter) disc-like structures that are
regularly rotating. At least some of these large discs represent
remnants of a major merger event that occurred at least $\sim$$10^9$ yr
ago (Morganti et al.\ 1997\nocite{1997AJ....113..937M}; Serra et al.\
2006\nocite{2006astro.ph..2621S}).  Our much deeper \HI\ study of the nearby
\sauron\ sample provides complementary information. Given the small
size of our sample, it is not surprising that such extremely \HI-rich
systems are not detected.  It is interesting to note, however, that,
while in the early-types detected by HIPASS the \HI\ is mainly
distributed in discs---consistent with the results of van Driel \& van
Woerden (1991\nocite{1991A&A...243...71V}), although the HIPASS
detects larger structures---the fainter \HI\ structures in the \sauron\
galaxies appear to have more varied morphologies. This may imply that
regular systems mainly occur in very gas-rich early-type galaxies.

From single-dish surveys it appears that peculiar galaxies have a
higher detection rate. However, Hibbard \& Sansom (2003\nocite{2003AJ....125..667H})
did not detect neutral hydrogen in early-type galaxies classified as
optically peculiar using the presence of `optical fine structure' (shell,
ripples, plumes etc.).  The failure to detect obvious tidal
\HI\ features (although with observations not as deep as those
presented here) suggests that if these fine-structures in early-type
galaxies are remnants of disc galaxy mergers, either the progenitors
were gas-poor and/or has been converted into other phases.
Our sample (and indeed the \sauron\/ parent sample) is formed by
regular galaxies where no major peculiarities have been observed (with
the possible exception of NGC\,2685). Nevertheless, we still have a
very high detection rate of \HI. This further strengthens the
conclusion that the relation between a galaxy being classified as peculiar and the
presence of neutral hydrogen is not straightforward.

Finally, it is worth noting that a shallower study of neutral hydrogen in a
sample of radio galaxies (Emonts et al.\ 2006a\nocite{2006astro.ph..8438E})
has shown that \HI\ emission is detected in 25\% of the
objects. Interestingly, the most \HI-rich structures are associated with
galaxies with compact (or small) radio sources, i.e., objects similar to
NGC\,4278 (see also Emonts et al.\ 2006b\nocite{eeta06}).

\section{Comparison with optical properties}
\label{sec:comparison_sauron}

We now consider the \HI\ properties in relation to the optical properties of
the galaxies. We concentrate on the observations made with \sauron, which
generally cover the central regions of the galaxies out to about an effective
radius, corresponding to roughly one or two times the WSRT beam size.

\begin{figure*}

\includegraphics[height=12.4cm]{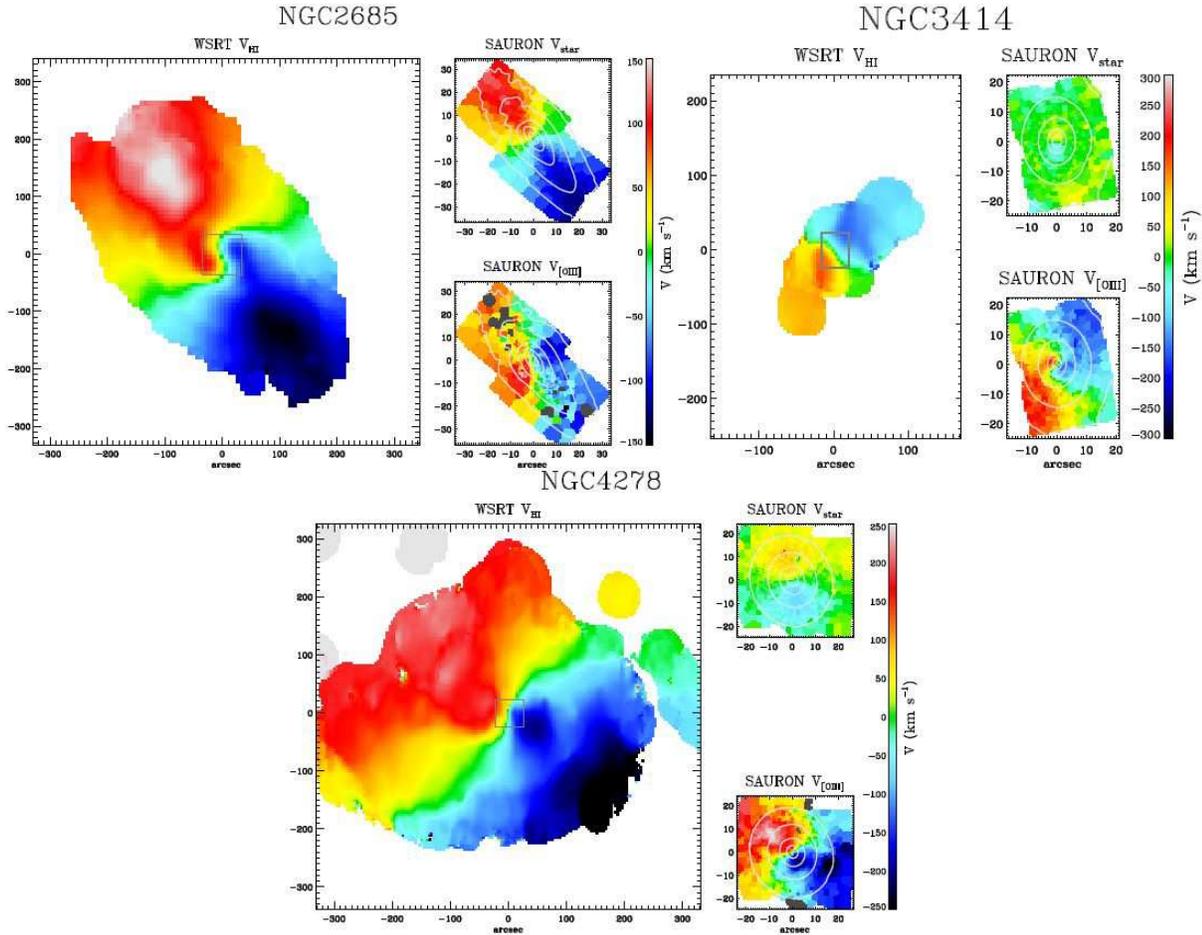}
\caption{Velocity fields of the neutral hydrogen, ionised
gas and stars for three of the four galaxies where regular \HI\ discs
have been detected: NGC\,2685, NGC\,3414 and NGC\,4278.\label{fig:neutral_ionised}}
\end{figure*}

\begin{figure*}
\includegraphics[width=13cm]{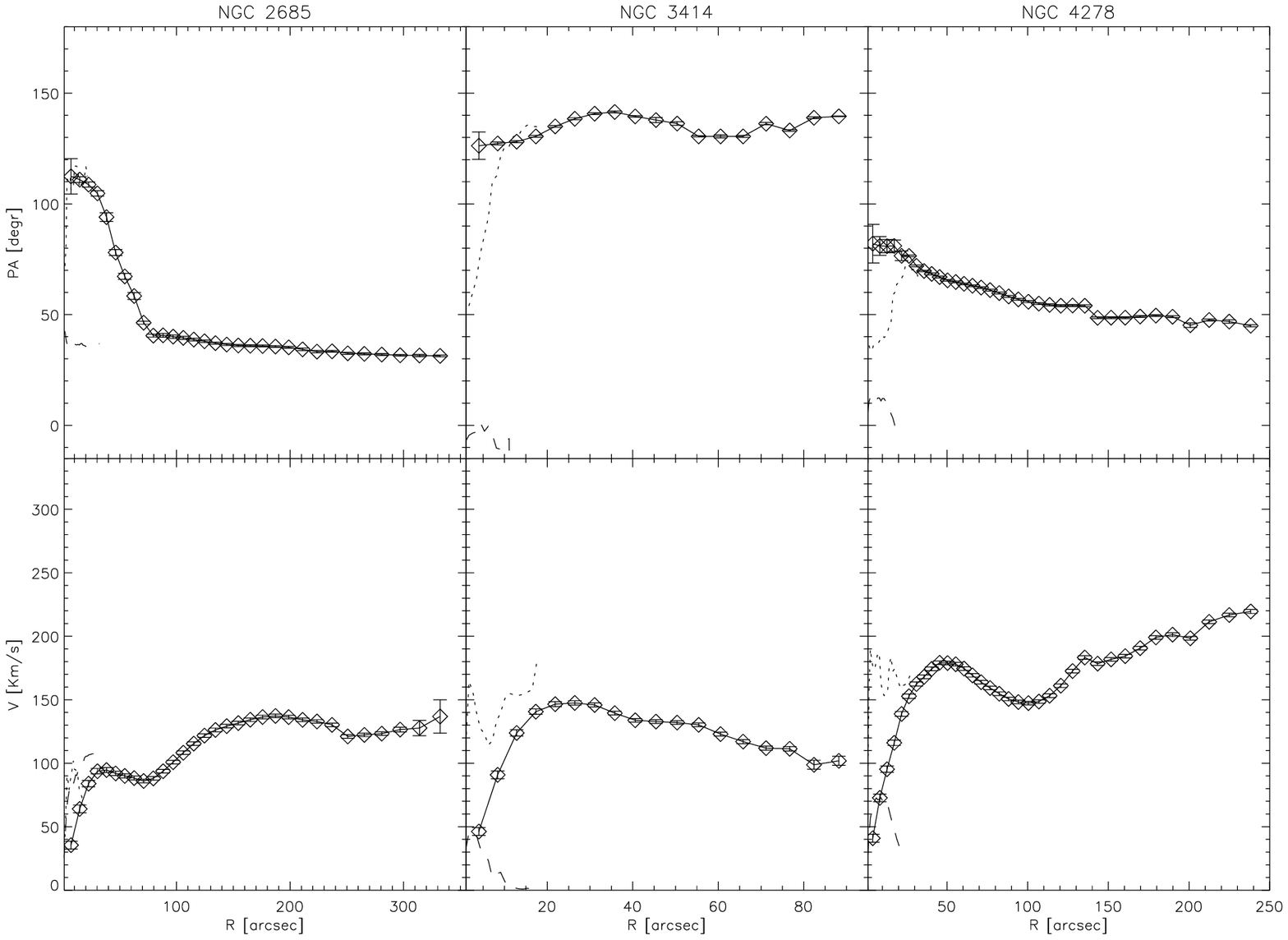}
\caption{Radial profiles of position angle (top) and  projected rotation
curves (bottom) of the neutral hydrogen in three galaxies that show regularly
rotating structures (open symbols). For comparison, we have over-plotted the
same quantities for the ionised gas (dotted lines) and stars (dashed lines) as
measured on the \sauron\ velocity maps. 
We subtracted 360\degr
from the position angle of the stellar component in NGC\,3414 for
presentation purposes. \label{fig:figelf}}

\end{figure*}

\subsection{Neutral hydrogen and the ionised gas}
\label{sec:comparison_sauron_ionised}

The WSRT observations suggest that $\sim$70\% of early-type galaxies in field
or in low-density environments have neutral hydrogen. For comparison, the
detection rate of ionised gas in the 48 \sauron\ (cluster and field) galaxies
is 75\% and it goes up to 83\% for field galaxies. Our observations show that
all the galaxies where some \HI\ has been detected also have ionised-gas
associated to them.

Although the linear scales studied with \sauron\ are very different from those
observed with the WSRT, the similar high occurrence of \HI\ and  ionised gas
suggests that most early-type galaxies have a detectable amount of gas and
that a link exists between the two phases (ionised and neutral) of the gas.
This is further motivated by the fact that earlier observations of a few
galaxies (using long-slit spectroscopy, e.g. NGC\,3108, J\'osza et al.\
2004b\nocite{2004IAUS..220..177J} and IC~4200 Serra et al.\
2006\nocite{2006astro.ph..2621S}) have found that, despite the different
scales involved, the kinematics of the ionised gas and the neutral gas are
very similar, suggesting that both gas phases are part of the same
structure. The combination of the WSRT data with the optical integral-field
spectroscopy available for the \sauron\ galaxies allows us to study this in
more detail.


The panels in Fig.~\ref{fig:neutral_ionised} show the velocity fields of the
\HI, of the ionised gas and of the stellar component for three of the four objects in
our sample with regular, disc-like \HI\ emission (NGC\,2685, NGC\,3414 and
NGC\,4278). We do not show the fourth galaxy where a regular
\HI\ disc has been found (NGC~4150) because the very weak \HI\ emission makes it difficult
to construct a meaningful velocity field.  

The velocity fields of Fig.~\ref{fig:neutral_ionised} show that, regardless of
the complicated character of the kinematics of the ionised gas in the very
central regions (in terms of, e.g., kinematical twists; see also Sarzi et al.\
2006\nocite{2006MNRAS.366.1151S}), the gas kinematics of the outer edges of
the regions covered by \sauron\ nicely match that of the corresponding very
inner data points of the \HI\ data.  We have traced the position angle and the
absolute rotation as function of radius using the harmonic-expansion method
for analysing two-dimensional velocity maps of Krajnovi\'c et al.\
(2006\nocite{2006MNRAS.366..787K}). The projected rotation curve is defined as $V(R)
= b_1(R)$, where $b_1$ is the amplitude of the cosine harmonic term, and $R$
is the length of the semi-major axis of the best fitting ellipse along which
velocities were extracted. $b_1$ is related to the circular velocity through
the inclination ($i$) of the \HI\ disk: $b_1 = V_{\rm rot}(R) \times \sin(i)$. The
ellipse position angle (PA) traces the orientation of the maximum rotation on the
maps. In Fig.~\ref{fig:figelf} we plot these two quantities for three galaxies
with detected large scale disc-like \HI\ distributions (for which such an
analysis is possible) and compare them with the same quantities measured on
the corresponding \sauron\ velocity maps (both of ionised gas and stars). The
\sauron\ data have higher resolution and show specific behaviour on small
scales, but on large scales nicely connect to the
\HI\ kinematics.
The agreement between the ionised and neutral gas is particularly good for the
PA's, but the rotation velocity amplitudes are also very consistent.  In all
three cases the stellar and gas PAs are strongly misaligned. NGC~2685 is a
somewhat special case because the orientation of the stellar velocity field
corresponds to the orientation of the large scale \HI\ disc, suggesting an
underlying connection between these two components, while the central warp
might be a more recent structure.  The rotation of the stellar component is
generally smaller (as expected) than the gas rotation.

A clear trend that is visible in our data is that {\sl galaxies with
regular \HI\ discs also have extended, kinematically regular
structures in the ionised gas}. Even NGC 2768 may fit this rule. In
this galaxy a large, very regular (polar) disc of ionised gas is seen
that does not have a regular neutral counterpart. Nevertheless, the
\HI\ in this galaxy seems to be an extension of the polar ionised
disc, and could be the outer remains of the accretion that lead to the
polar structure, so a relation between the \HI\ and the ionised gas
may exist. Conversely, the galaxies with `offset \HI\ clouds' have
fainter emission from the ionised gas, while also this ionised gas
seems to form a less regular structure.  In NGC\,1023, the galaxy with
a large, less regular \HI\ structure, the kinematics of the ionised
gas is very similar to that of the neutral gas.  In two (NGC\,2549 and
NGC\,7457) of the three objects without \HI\ detection, there is
evidence of \Oiii\ and \Hb\ emission, but this emission is very faint
compared to the other cases mentioned above. NGC\,5308 is the only
case completely devoid of ionised gas and also no \HI\ has been
found. In NGC\,5982 ionised gas is not visible in the direction of the
blob (or on that side of the galaxy). Ionised gas emission is
localised on the central 10$^{\prime\prime}$ with a tail going
South and the gas velocities are consistent with the redshift of the
\HI\ cloud.

\subsection{Nature of the host galaxy}
\label{sec:comparison_sauron_host}

Our observations have revealed \HI\ in all four galaxies classified as
E in the RC3, and in five out of the eight S0s. This is perhaps
surprising, as one might have naively expected that the presence of an
\HI\ disc is connected with a stellar disc, and that \HI\ therefore
would be detected more often in the S0 galaxies than in the true
elliptical galaxies. This is clearly not the case. Moreover, one of
the extended \HI\ discs is found in the E galaxy NGC\,4278.

The \sauron\ stellar kinematics show that early-type galaxies can be
classified more physically as slow- and fast-rotators, based on a measure of
their specific angular momentum (Emsellem et al. in prep.), with the former
being fairly isotropic and mildly triaxial objects, and the latter nearly
axisymmetric and radially anisotropic (Cappellari et al.\
2005).  Nine of our objects are fast-rotators, and only three
are slow-rotators {(NGC~3414, NGC~5198 and NGC~5982)}.

\begin{figure}
\plotone{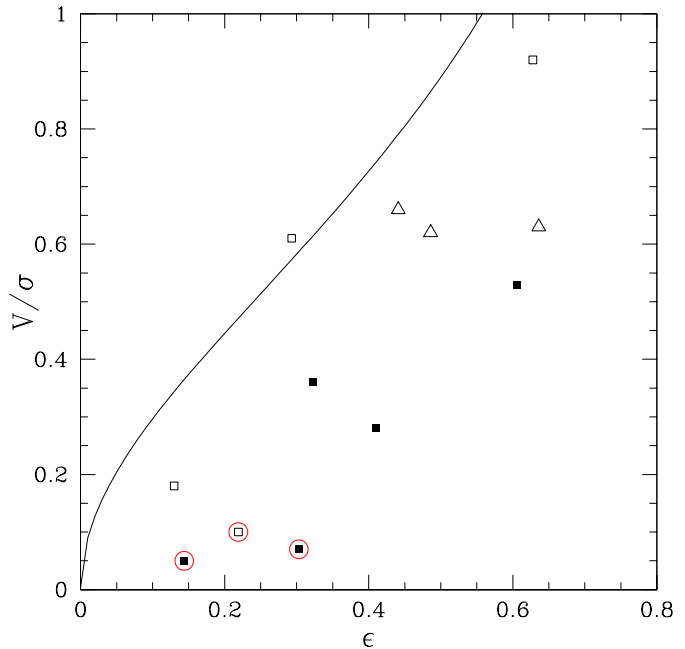}
\caption{Ratio of ordered over random motion $V/\sigma$ versus the
observed flattening $\epsilon$ for the 12 objects in the sample
studied in this paper  (values taken from Cappellari et al. in prep.)}. The
oblate isotropic rotator line (Binney 2005) is also shown. The \HI\
morphology of each object is indicated
by different symbols: triangle: \HI\ upper limits, filled squares:
\HI\ offset blobs or messy, open squares: \HI\ discs. Symbols with a circle
indicate the three  slow rotators. \label{fig:figstellarkin}.
\end{figure}

Fig.~\ref{fig:figstellarkin} shows the classical $V/\sigma$ versus $\epsilon$
diagram for our twelve objects.  The data show no trend between the
detection and/or the morphology of the \HI\ and the dynamical type of
galaxy. In particular, \HI\ is detected (including one
\HI\ disc) in all slow-rotators while all the \HI\ upper limits are found for
fast-rotators. More generally, \HI\ detections are uniformly spread through
the $(V/\sigma, \epsilon)$ diagram.  If fast and slow rotators represent the
relics of different formation paths, this does not appear in the presence and
distribution of the \HI.

\subsection{Neutral hydrogen and stellar population}
\label{sec:comparison_sauron_pops}

Given the evidence linking the neutral gas to ionised gas components and to
evidence of accretion, it is natural to expect a link between the neutral gas
content and episodes of star formation. We therefore investigate whether the
luminosity-weighted age of the stellar population derived from the \sauron\
data (Kuntschner et al.\ 2006\nocite{keta06} and Kuntschner et al. in prep.), within 1$R_{\rm e}$, using the
population models by Thomas, Maraston \& Bender (2003), correlates with the
presence or morphology of the neutral hydrogen.

We find that no clear trend emerges: a relatively young stellar
component can be found both in galaxies detected in \HI\
(e.g.\ NGC\,4150) as well as in galaxies undetected (e.g.\
NGC\,7457). Moreover, \HI\ detected galaxies can be dominated by an
old stellar population (e.g.\ NGC\,1023, NGC\,3414, NGC\,4278). In
particular for NGC\,1023 this is remarkable as the complex kinematics
of the \HI\ in this galaxy suggests that significant accretion of
gas has occurred recently.

\begin{figure}
\plotone{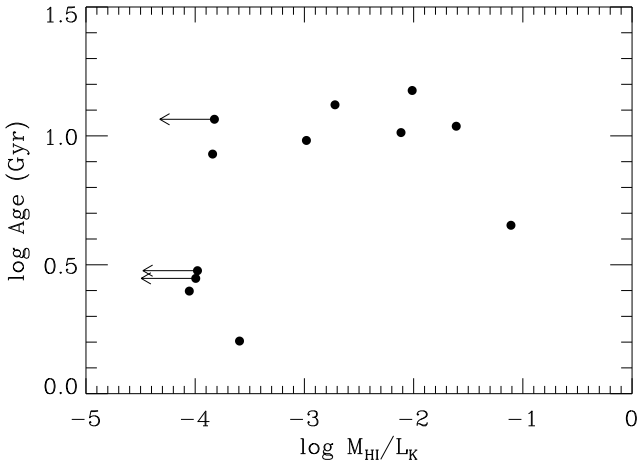}
\caption{Estimates of the luminosity weighted stellar age, as derived from the
\sauron\ data within 1$R_e$  plotted against the \HI\ mass fraction
($M_{\HI}/L_K$). 
Arrows indicate upper limit to the \HI\ mass fraction in galaxies with no gas detection.
We use the $K$-band luminosity, instead of the $B$-band, as it is much less
sensitive to dust absorption effects, and because it better traces the bulk of
the dominant old stellar population of early-type
galaxies. \label{fig:figmassfraction}}
\end{figure}

Fig.~\ref{fig:figmassfraction} illustrates the distribution of the global
stellar age versus the \HI\ mass fraction ($M_{\rm HI}$/$L_K$).  We use the
K-band luminosity, instead of the B-band, as it is much less sensitive to dust
absorption effects, and because it better traces the bulk of the dominant old
stellar population of early-type galaxies Although the statistics is extremely
limited, it is intriguing to note that galaxies with  a relatively
young  stellar population are found among those with a low \HI\
mass-fraction. The galaxies with high \HI\ mass-fraction tend to have an old
stellar component.

The typical column densities of the \HI\ (at the modest spatial resolution of
our observations) are relatively low, at most a few times $10^{20}$ cm$^{-2}$.
This, therefore, confirms what was already found in other \HI\ studies of
early-type galaxies (e.g. van Driel \& van Woerden
1991\nocite{1991A&A...243...71V}; Morganti et al.\
1997\nocite{1997AJ....113..937M}), namely that even in the cases where a large
reservoir of \HI\ is found, the gas is spread over a large area and therefore
very diluted and not able to reach, at least on large scales, a column density
high enough for star formation to occur (Kennicutt
1989\nocite{1989ApJ...344..685K}; Schaye
2004\nocite{2004ApJ...609..667S}). This result has been taken as an indication
that a large \HI\ reservoir can stay around for a very long time without being
consumed by any star forming activity. This could mean that many systems
acquire gas but that only in some this gas manages to form stars (see
Fig.~\ref{fig:figmassfraction}) and gets consumed while in others this is not
the case. Hence, we do not find any correlation between the
presence/morphology of the
\HI\ and the stellar population.

However, it is worth noting that individual cases do exist where the merger
origin of the galaxy has been confirmed by the similarities between the age of
the \HI\ structures and the age of the youngest of the stellar population
detected. In the \HI-rich early-type galaxies IC~4200 and B2~0648+27 (Serra et
al.\ 2006\nocite{2006astro.ph..2621S}; Emonts et al.\ 2006b\nocite{eeta06},
respectively) the age of the youngest stellar population is found to be about
2~Gyr for IC~4200 and 0.3 Gyr for B2~0648+27 while the age of the \HI\
structure derived from the regularity of the \HI\ distribution (i.e., the
neutral hydrogen must have had the time to complete 1--2 orbits) is
(1--2~Gyr)1--2~Gyr.  Thus, both diagnostics suggest that both these galaxies
have formed between 1 and 3 Gyr ago as a result of major mergers.

Another possibility is that the neutral hydrogen comes from smooth, cool
accretion of the IGM (see, e.g., Keres et al.\
2005\nocite{2005MNRAS.363....2K} and refs therein): in this case  we do
not expect a major star formation event that would leave a distinct marker in
the stellar population as would be the case in a merger, but perhaps more
continuous small episodes of star formation from the smoothly acquired gas.
Therefore, a correlation between the \HI\ amount/structures and the age of the
stellar population would not necessarily be expected since the young stellar
population would be small. This appears to be a a realistic possibility for
some of the galaxies in our sample (see also Section~\ref{sec:origin}).

\section {\HI\ and radio AGN}
\label{sec:comparison_agn}

The  nuclear activity in galaxies is often being explained as triggered by
merger and/or interaction processes.  Torques and shocks during the merger can
remove angular momentum from the gas in the merging galaxies and this provides
injection of substantial amounts of gas/dust into the central nuclear regions
(see, e.g., Mihos \& Hernquist 1996\nocite{1996ApJ...464..641M}). This can
lead to kinematically distinct components. The
presence of gas associated with most of the galaxies in our sample and the
presence of kinematically complex stellar features (e.g., KDCs) suggest that
at least some of the galaxies have experienced an interaction/merger in their
past, and makes it interesting to explore whether any relation exists between
the observed characteristics of the gas and the presence of nuclear radio
emission that could be the result of an active nucleus.

We have derived for each galaxy the flux of the radio continuum (or upper
limit to it) from the images obtained from the line-free channels of our line
observations. In eight of the 12 galaxies the radio continuum is detected and
it appears in all cases unresolved and coincident with the optical centre.
The radio powers are listed in Table~\ref{tab:table2}. The galaxies are (with
one exception, NGC\,4278) all low-power radio sources (i.e., $<10^{21}$ W
Hz$^{-1}$, well below the typical power of radio galaxies). At these levels of
emission, the radio continuum can be both due to an active nucleus or to star
formation in the nuclear regions. Based on the flat spectrum typical of the
radio component, a study at 3.6 cm of nearby E/S0 galaxies (with four objects
in common with our sample) identified low-power AGN as the most likely source
in a significant fraction of the objects (Krajnovi\'c \& Jaffe
2002\nocite{2002A&A...390..423K}). In addition to this, NGC\,4278 is a
well-known weak radio galaxy where radio jets have been observed (Giroletti,
Taylor \& Giovannini 2005\nocite{2005ApJ...622..178G} and references therein)
and where the radio emission is known to originate from the AGN. It is also
worth noting that in most of the galaxies of our sample, the ionised gas
appears to be excited by sources other than O-stars (given the high \Oiii/\Hb\
ratios), and that only in a minority of S0 galaxies has on-going star
formation been observed (see also Section~\ref{sec:comparison_sauron} and
Sarzi et al.\ 2006\nocite{2006MNRAS.366.1151S}).  It is, however, still
possible that different physical mechanisms could account for the weak radio
continuum sources in E and S0 galaxies (e.g.,  Wrobel \& Heeschen 1991).

Interestingly, Table~\ref{tab:table1} shows that the galaxies undetected in
\HI\ are also undetected in radio continuum (note that NGC\,7332 is indicated
as detection in \HI\ but it is in fact an uncertain case as discussed in
Section~\ref{sec:results_offset}).  The presence of a possible connection
between the two (to be confirmed with a larger sample) suggests that whatever
the process is that brings the neutral gas in/around these objects, it can
also lead to the triggering of some activity in the nucleus.

\section{Origin of the neutral gas}
\label{sec:origin}

The similarities in the kinematics of the neutral and the ionised gas
suggest that they are simply two phases of the same structure, which
share the same origin. Based on the distribution of the kinematic
misaligment between ionised gas and stars within the full \sauron\/
sample of E and S0s, Sarzi et al.\ (2006\nocite{2006MNRAS.366.1151S})
conclude that the gas cannot be all internal or all external.  For
fast-rotating galaxies, there is a higher incidence of co-rotating gas
and stars, suggesting that the two are closely linked, favouring an
internal origin in some cases. For slow-rotating galaxies, which tend
to be rounder and more triaxial, the kinematic orientations of the
stars and gas have no preferred alignment, suggesting that external
accretion can explain the presence of gas in these systems.

Our \HI\ data show, in agreement with other studies (e.g., Knapp et al.\
1985\nocite{1985AJ.....90..454K}), that the distribution of the ratio $M_{\rm
HI}$/$L_B$ derived from our observations is very broad and the \HI\ content is
uncorrelated with the optical luminosity of the galaxy (unlike the case of
spiral galaxies).  This can be taken as an indication of an external origin
for the gas. In addition to this argument, in a large fraction of the objects
discussed in this paper, the distribution and kinematics of the \HI\ (i.e.\
off-set clouds/polar discs) clearly indicate an external origin. It appears,
therefore, that at least small accretion events are very common in the life of
every early-type galaxy.  Indeed, although an internal origin of the gas
cannot be completely ruled out, it is difficult to identify any of our
galaxies where the origin of the \HI\ could be {\sl completely} internal
(i.e., the result of stellar mass loss). Even in the case of NGC\,4150 where
most of the \HI\ comes from a very small (galaxy-scale) disc that, in
principle, could have an internal origin, very faint \HI\ emission is also
detected well outside the galaxy, while also the stellar kinematics show the
presence of a small kinematically decoupled core (composed of relatively young
stars) that seems to indicate that a merger/accretion event must have occurred
in the recent past of this galaxy.

While mergers and small-companion accretions appear to be at the origin of
some of the observed \HI\ structures it is,
however, confusing to see that the stellar population does not reflect the
presence of a younger component related to a recent merger/accretion. An
extreme case is NGC\,1023 where, despite the large amount of neutral hydrogen
distributed in a complex and very clumpy way perhaps as a result of a recent interaction, only
an old stellar population is observed. In other cases with more regular
structures,  the \HI\ column
densities are low so that little star formation has occured and will occur so
that  the star
formation triggered by the merger is a small fraction of the stellar mass and
therefore difficult to detect.  In most of the galaxies in our sample, the
evidence, from the optical point of view, of a recent merger/interaction are
quite subtle (except for NGC\,2685). This may again suggest that the neutral
hydrogen is able to stay around (given suitable conditions of the environment)
for long periods without leading to obvious optical features. It can also indicate that instead of a merger event, the
\HI\ originates from smooth, cold accretion of IGM. The possibility of forming
even polar ring structures in this way has been described in Macci\'o, Moore
\& Stadel (2006\nocite{2006ApJ...636..25M}). Indeed, this could be supported,
at least for some cases, from the lack of relation between the presence of the
\HI\ and the age of the stellar population.

While steady cold accretion may work on large scales, it is not clear whether
a similar mechanism can also explain the presence of central stellar decoupled
components as well as the possible connection between the presence of central
radio continuum and the presence of neutral hydrogen
(Section~\ref{sec:comparison_agn}). If this relation is confirmed by larger
samples, a way to explain it is via a merger/accretion event that not only can
supply the large-scale gas, but that is also able to supply a fraction of the
gas to the nuclear regions.  The typical time-scale of the radio emission
($10^7 - 10^8$ yr) is much shorter than the time-scale the \HI\ can stay
around (see, e.g., Section~\ref{sec:comparison_sauron_host}), therefore either
these low-power AGN go through multiple periods of activity or the activity is
triggered at a later stage of the merger (as it seems to be the case for more
powerful radio galaxies, see, e.g., Tadhunter et al.\
2005\nocite{MNRAS.356..480T}; Emonts et al.\ 2006b\nocite{eeta06}).

\section{Conclusions}
\label{sec:conclusions}

The main result of these observations is that, in terms of the neutral gas,
{\sl the class of early-type galaxies is extremely rich and varied and that
galaxies that {\it prima facie} appear very similar show subtle but important
differences}. Our detection rate of \HI\ ($\sim 70$\%) is comparable to that
ionised gas (75\%, Sarzi et al.\ 2006\nocite{2005astro.ph.11307S}). This is
surprisingly high compared with earlier studies, likely the consequence of the
fact that our \ HI\/ data are a factor 50-100 deeper. As with the ionised gas,
we find a variety of structures, including irregular distributions (likely the
remnant of a recent accretion), small polar rings, strongly-warped (up to
90$^\circ$) structures, and extended re gular discs (in some cases containing
as much \HI\/ as the Milky Way!). The peak column density in these discs is
low - below (and sometimes well below) a few times $10^{20}$ cm$^{-2}$ - so no
significant star formation is occurring, implying that these d isks may
survive for a very long time.

A very clear correlation exists between the presence and the properties of
ionised and neutral gas. They share the same kinematics, showing that they are
two phases of the same structure. On the other hand, the correlation of the
gas kinematics (neutral and ionised) with that of the stars is quite
poor. Most interestingly, we find that neither the amount of gas, nor its
kinematics, correlate clearly with the stellar dynamics: the galaxies with
slow stellar rotation (likely to be `true' elliptical galaxies) are all
detected in \HI, with some containing regular gas discs. Conversely, many of
the fast-rotating galaxies (likely S0-like galaxies) are not detected. This
range of gas properties does not intuitively reflect the structural
differences between elliptical and S0 galaxies, but suggests that the
relationship between the gaseous and stellar components is more complex than
previously thought.

Moreover, the amount of neutral gas {\sl does not} appear to correlate with the stellar
population characteristics. In a few galaxies, large amounts of \HI\ are
detected while the luminosity-weighted stellar population is purely old,
showing no evidence of young stars that may have formed from the
gas. Conversely, there are galaxies with a young luminosity-weighted age where
no gas is detected. This is contrary to the commonly held idea that accretion
and merging trigger star formation in the central regions. This suggests that
other modes of accretion also exist, and may hint that `cold accretion' does
occur in some systems.

In the galaxies detected in \HI\ we also detect radio  continuum radiation
which may be from a small AGN or result from star formation. In the \HI\
non-detections, no such source is detected. This suggests that whatever
process brings the neutral gas to early-type galaxies, it can trigger some
activity/star formation in the central regions.

Our survey indicates that the presence of neutral gas is common in field
early-type galaxies, if sufficiently deep observations are available. This is
further supported by the fact that the observed galaxies were not selected
based on any peculiarity and are relatively regular objects.  Our results
point to an external origin of the gas, suggesting that gas accretion is
common and does not happen only in peculiar early-type galaxies. This may be
connected to the high incidence of small-scale kinematical features in the
\sauron\ data.
This accretion can happen in many different ways, and can
involve a very large range of masses. Also in
spiral galaxies, gas accretion events are found to be very common over their
lifetime (Naab \& Ostriker 2006\nocite{2006MNRAS.366..899N}, van der Hulst \&
Sancisi 2004). From our study it appears that the distribution, more than the
amount, of \HI\ is an important element in determining the morphology of the
galaxy. Even in the very \HI-rich early-type galaxies, the neutral hydrogen is
always distributed over very large areas therefore it has always a very low
column density, too low for star formation to occur.

Our current sample includes only 12 galaxies restricted to low density
environments, and the statistical validity of our results is therefore
limited.  Given the links that seem to be present between the characteristics
of the neutral and ionised gas components, it is important to re-examine also
the cluster environment to fully explore the effect/importance that the
environment has on our results. In cluster galaxies, the fraction of ionised
gas detections --- derived from the
\sauron\ study --- is about 55\%, and therefore, if the relation
holds, a significant number of galaxies may indeed show neutral
hydrogen even in this more hostile environment.

\section*{Acknowledgements}

We are grateful to Gyula J{\'o}zsa for providing the data-cube of
NGC\,2685. We thank Harald Kuntschner for providing the stellar population
ages and Eric Emseller for providing the galaxy classification. We thank the
referee, Elaine Sadler, for useful comments which improved the paper. This
research was supported by the Netherlands Research School for Astronomy NOVA,
and by the Netherlands Organization for Scientific Research (NWO) through
grant 614.000.426 (to AW) and VENI grant 639.041.203 (to MC). The Westerbork
Synthesis Radio Telescope is operated by the Netherlands Foundation for
Research in Astronomy ASTRON, with support of NWO.  This research has made use
of the NASA/IPAC Extragalactic Database (NED) which is operated by the Jet
Propulsion Laboratory, California Institute of Technology, under contract with
the National Aeronautics and Space Administration. The Digitized Sky Survey
was produced at the Space Telescope Science Institute under US Government
grant NAG W-2166.

\appendix

\section[]{Serendipitous discovery of a possible OH megamaser}
\label{appendix:a}

\begin{figure}
\includegraphics[width=6.3cm,angle=-90]{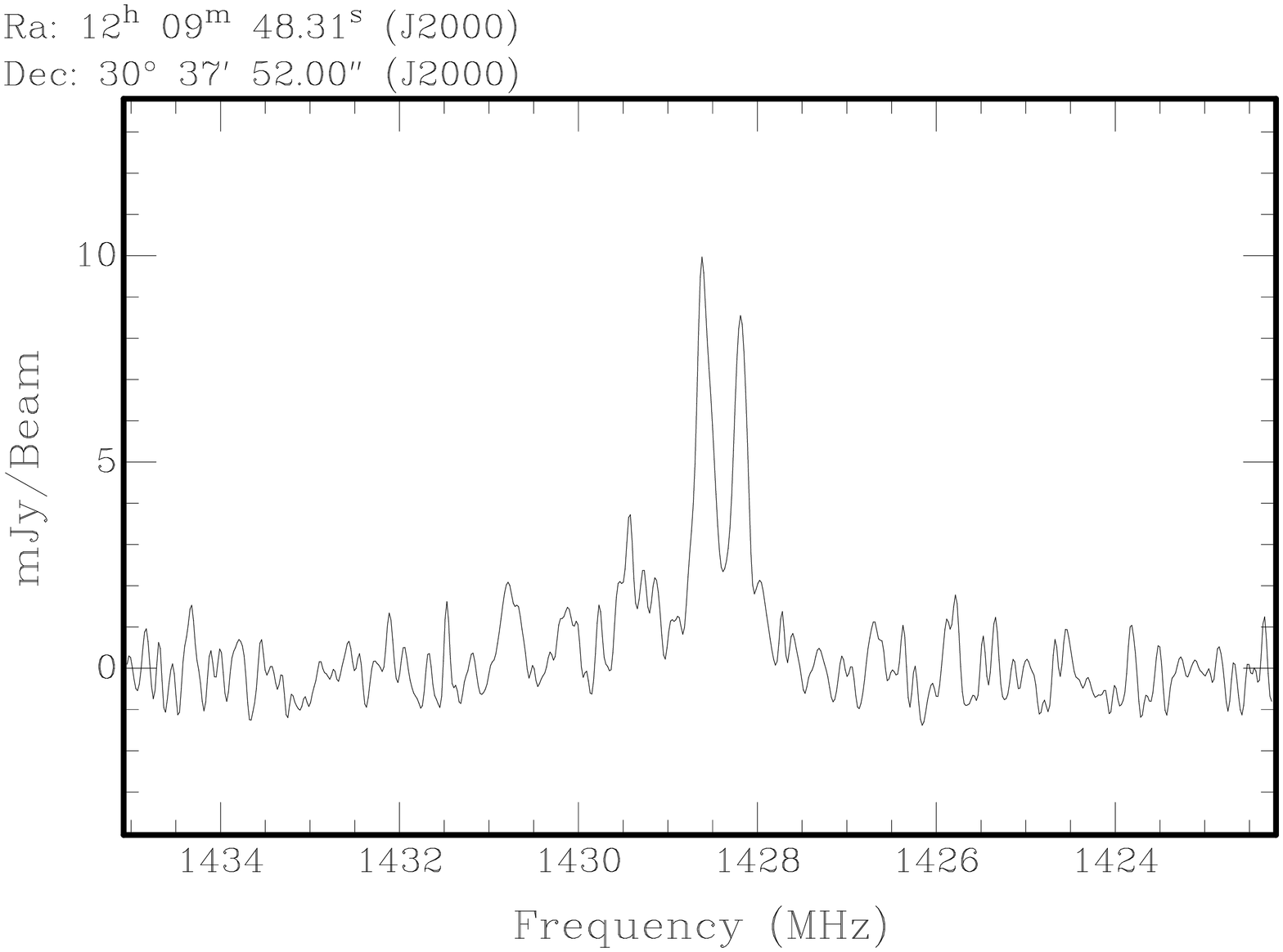}
\caption{Spectrum of the serendipitous emission in the field of NGC~4150.\label{fig:figaone}}
\end{figure}

All the \HI\ observations presented in this paper were carried out using the
wide band (20 MHz) offered by the WSRT. This means that emission can be
detected over a wide range of velocities.  In the observations of NGC\,4150,
we have detected an emission located more than $10^{\prime}$ from the
centre of the field (i.e. NGC~4150). The observed complex spectrum is shown in
Figure~\ref{fig:figaone}.  Two main peaks are observed (at the frequencies of
$\sim 1428.24$ and 1428.63 MHz), but the emission covers a range of at least
$\sim $ 3~MHz.
If the emission would originate from neutral hydrogen, it would correspond
to a systemic velocity of $-1700$ \kms.  A more realistic possibility is that
the detection corresponds to an emission line different from \HI.

\begin{figure}
\plotone{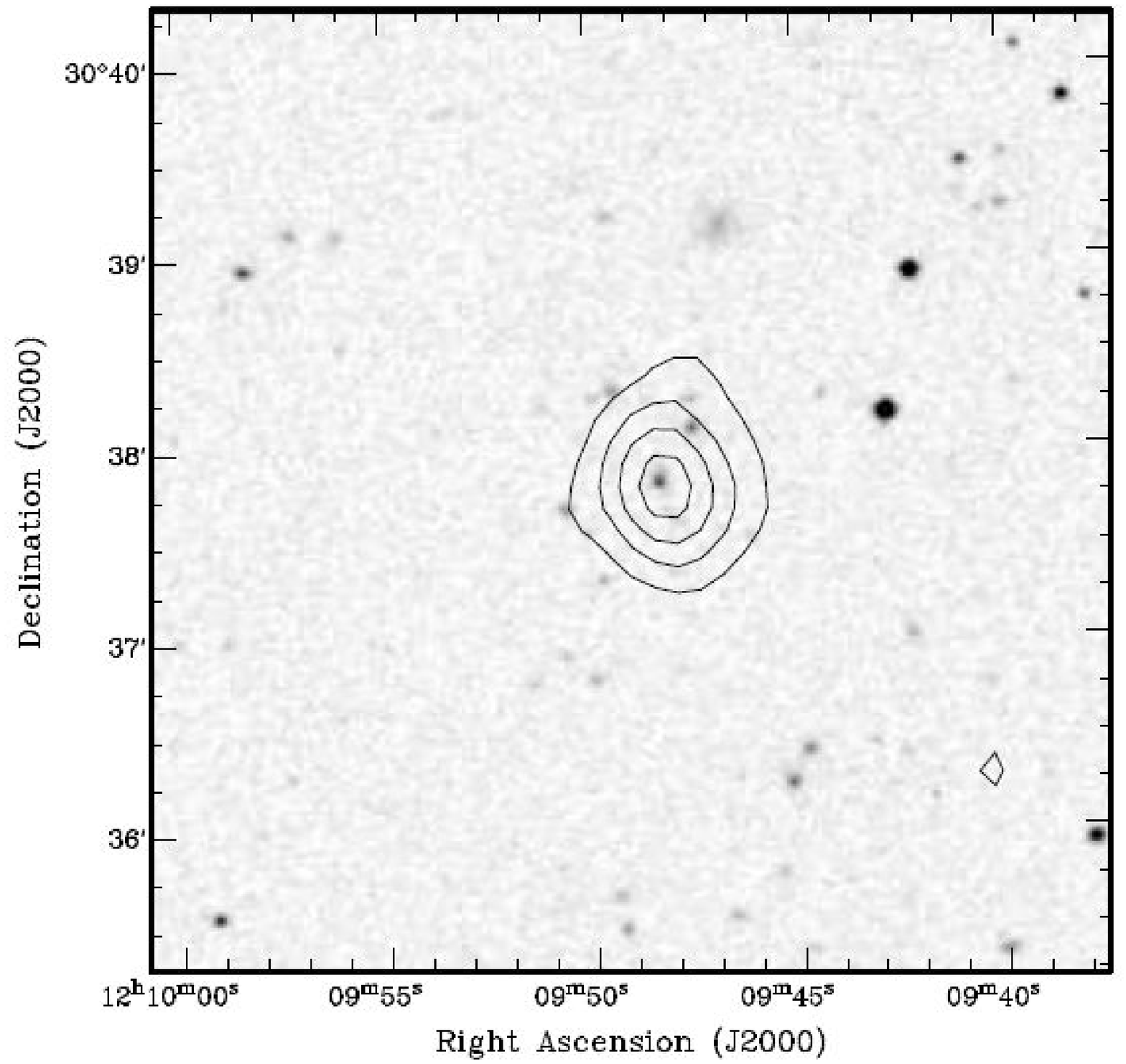}
\caption{ Total intensity of the possible detection of an OH megamaser
          in the field of NGC\,4150.\label{fig:figatwo}}
\end{figure}

The total intensity of the emission superimposed onto an optical image is
shown in Figure~\ref{fig:figatwo}. As visible in the figure, the emission
(located at R.A.\ $= 12^{\rm h} 09^{\rm m} 48^{\rm s}$ and $\delta = 30^\circ
37^\prime 52^{\prime\prime}$ J2000) has an optical counterpart, also identify
as the IRAS source IRAS F12072+3054.  Continuum emission is also detected at
the location of the line emission. The flux of the continuum is 270 $\mu$Jy.

We suggest that the detected emission could originate from an OH megamaser,
often found in IR bright galaxies (see e.g. Baan, Salzer \& Lewinter 1998).
The main lines of the hydroxyl emission are at 1665 and 1667 MHz.  This would
mean that the emitting galaxy is located at $z \sim 0.17$. If confirmed, this
would represent the first OH megamaser discovered serendipitously with the
WSRT.

\label{lastpage}

\end{document}